\documentclass[openacc]{rstransa}

\usepackage{bbold}
\usepackage{graphicx}
\usepackage{pdfpages}
\usepackage{adjustbox}
\usepackage{booktabs}
\usepackage{makecell}
\usepackage{algorithm,algpseudocode}
\usepackage[capitalise, noabbrev]{cleveref}
\usepackage{mathtools}
\usepackage{url}
\usepackage{fixltx2e}
\usepackage{enumitem}
\usepackage{etoolbox}

\usepackage{subcaption}

\graphicspath{{./Figures/}}

\algnewcommand{\Inputs}[1]{%
  \State \textbf{Inputs:}
  \Statex \hspace*{\algorithmicindent}\parbox[t]{.8\linewidth}{\raggedright #1}
}
\algnewcommand{\Initialize}[1]{%
  \State \textbf{Initialization:}
  \Statex \hspace*{\algorithmicindent}\parbox[t]{.8\linewidth}{\raggedright #1}
}

\begin{document}
\title{Generative Priors for MRI Reconstruction Trained from Magnitude-Only Images Using Phase Augmentation} 
\author{
Guanxiong Luo$^{1,4}$, Xiaoqing Wang$^{5}$, Mortiz Blumenthal$^{1,2}$, Martin Schilling$^{1}$, Erik Hans Ulrich Rauf$^3$, Raviteja Kotikalapudi$^3$, Niels K Focke$^3$, Martin Uecker$^{1,2,4}$}

\address{$^{1}$Institute for Diagnostic and Interventional Radiology, University Medical Center G\"ottingen, Germany\\
$^{2}$Institute of Biomedical Imaging, Graz University of Technology, Graz, Austria\\
$^{3}$Clinic for Neurology, University Medical Center Göttingen, Germany\\
$^{4}$German Centre for Cardiovascular Research (DZHK), partner site Lower Saxony, Germany\\
$^{5}$Martinos Center for Biomedical Imaging, Massachusetts General Hospital, Harvard Medical School
}

\subject{medical imaging, generative models, inverse problem}


\corres{Guanxiong Luo, University Medical Center G\"ottingen, Institute for Diagnostic and Interventional Radiology, Robert-Koch-Str. 40, 37075 G\"ottingen, Germany.\\
\email{guanxiong.luo@med.uni-goettingen.de}}

\begin{abstract}
    In this work, we present a workflow to construct generic and  
    robust generative image priors from magnitude-only images. The priors can then 
    be used for regularization in reconstruction to improve image quality.
    The workflow begins with the preparation of 
    training datasets from magnitude-only MR images. This dataset is then
    augmented with phase information and used to train generative priors
    of complex images. Finally, trained priors are evaluated using
    both linear and nonlinear reconstruction for compressed sensing
    parallel imaging with various undersampling schemes.
    The results of our experiments demonstrate that
    priors trained on complex images outperform priors trained only
    on magnitude images. Additionally, a prior trained on a larger
    dataset exhibits higher robustness. Finally, we show that the
    generative priors are superior to $\ell^1$-wavelet regularization for
    compressed sensing parallel imaging with high undersampling.
    These findings stress the importance of incorporating phase
    information and leveraging large datasets to raise the
    performance and reliability of the generative priors for 
    MRI reconstruction. Phase augmentation makes it possible to
    use existing image databases for training.
\end{abstract}

\def\authorfont{\fontfamily{\sfdefault}\fontseries{m}\fontshape{n}\fontsize{10}{17}\selectfont\raggedright\color{jobcolor}}
\def\addressfont{\fontfamily{\sfdefault}\fontseries{m}\fontshape{n}\fontsize{10}{12}\selectfont\raggedright}

\maketitle
\section{Introduction}
Magnetic resonance imaging (MRI) is a widely used non-invasive technique, but a key challenge lies in balancing imaging speed with image quality. This compromise is primarily determined by the k-space measurements, which trace defined trajectories in the spatial Fourier domain.
Regularizing the inverse problem for parallel MRI reconstruction is 
an effective and flexible approach for improving image quality,
especially when the obtained k-space is highly undersampled in order
to shorten the scan time. The prior knowledge that images are
sparse in a transform domain as used in compressed sensing is known 
as $\ell^1$-norm regularization\cite{Lustig_Magn.Reson.Med._2007,Block_Magn.Reson.Med._2007}.
Combined with incoherent sampling, this allows recovery of images
from moderately undersampled k-space data with clinically acceptable
quality\cite{Vasanawala_Radiol._2010, Feng_J.Magn.Reson.Imaging_2017}. 

The application of deep learning makes it possible to further
increase undersampling without compromising image quality by 
leveraging the learned prior information from a training dataset.
Popular methods can be classified into three main categories: 
supervised methods\cite{Yang_NIPS_2016,Hammernik_Magn.Reson.Med._2017},
where the neural network is a result of unrolling an iterative
algorithm trained with labels and used to predict the reconstruction,
self-supervised methods\cite{Yaman_Magn.Reson.Med._2020, Blumenthal__2022}
that involve splitting the acquired k-space data of a scan
into two disjoint sets where only the first set is used
for reconstruction and the second set provides supervision,
and decoupled methods\cite{Tezcan_IEEETrans.Med.Imag._2018,Liu_Magn.Reson.Med._2020,Luo_Magn.Reson.Med._2020},
where a generative model or a denoiser is trained
to learn the empirical distribution of data which is
then used in a conventional iterative reconstruction
method. In the following, we will refer to a generative 
model also as a prior.

Training in supervised methods based on unrolled iterative
algorithms requires not only fully sampled k-space data, but
also pre-defined sampling patterns and precomputed coil sensitivities.
The prior knowledge learned in this way then pertains to these pre-defined 
settings. However, protocol settings for clinical and research are
changed often, and the preparation of reference data, which is 
used as labels for training, is costly.
Decoupled methods are able to avoid these constraints, and
the learned prior can even be transferred to new scenarios
such as different contrasts\cite{Luo_Magn.Reson.Med._2023}.

As a crucial part of decoupled methods, the use of generative models,
such as variational autoencoders and autoregressive models, was
investigated previously by formulating the linear reconstruction
problem for accelerated MRI from the Bayesian perspective and solving
it via maximum a posterior (MAP) estimation
\cite{Tezcan_IEEETrans.Med.Imag._2018,Luo_Magn.Reson.Med._2020}.
More recently, diffusion models emerged as effective
priors for MRI reconstruction and were combined with Monte Carlo
methods that sampling the posterior\cite{Jalal_NIPS_2021, Chung_Med.Image.Anal._2022, Luo_Magn.Reson.Med._2023}.
However, their performance is heavily dependent on the size and
quality of the training dataset and the computational resources available.

\begin{figure}
  \centering
  \includegraphics[width=0.5\columnwidth]{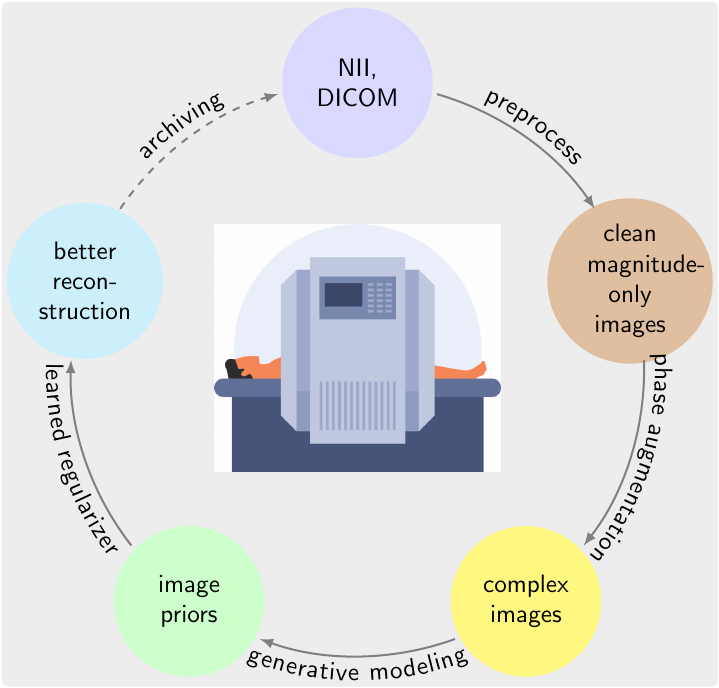}
  \caption{
The proposed workflow for extracting prior knowledge and
  using it for regularization in image reconstruction.
It comprises data preparation, phase augmentation, generative modeling, and concludes with the
use as learned regularizers in reconstruction. \label{fig:overview}}
\end{figure}

For this reason, it is desirable to use existing databases of MR images
for training. But as shown here, training from magnitude-only images leads
to inferior priors. This work therefore proposes a new approach
to construct priors using magnitude-only training images as illustrated 
in \cref{fig:overview}. The workflow begins with the preparation of training
datasets from magnitude-only MR images.
This dataset is then augmented with phase information and used to train
generative priors on complex images. Finally, trained priors can be
used with both linear and nonlinear reconstruction for
compressed sensing parallel imaging. The contributions of our work are:
\begin{enumerate}[wide,labelindent=0pt]
  \item[\textbf{Complex vs. magnitude-only priors:}] We demonstrate
that priors trained on complex images are superior to priors trained
only on magnitude images.
  \item[\textbf{Phase augmentation:}] We leverage a diffusion model trained on a
small dataset (\textasciitilde 1000 images) of complex images to augment a
much larger dataset (\textasciitilde 80k images) for which the phase
information of the image is not available.
  \item[\textbf{Robustness:}] We show that we can train more robust
generative priors by incorporating knowledge from a larger training dataset,
which contains a diverse range of images. Furthermore, the robustness
are verified in different reconstruction scenarios that involve different
sampling patterns, changes in TE/TR, different scanners and so on.
Such a database can be obtained by phase augmentation of magnitude images
which are readily available. 
  \item[\textbf{Flexibility:}] By integrating the priors as regularization
terms into existing reconstruction techniques, we maintain the flexibility
of existing reconstruction algorithms (linear and nonlinear) that can be used with various 
undersampling schemes and receive coils.
\end{enumerate} 
Parts of this work have been presented in Refs. \cite{Luo__2022a, Luo__2021}.

\section{Theory}

\subsection{Linear and nonlinear reconstruction}

The reconstruction in parallel imaging can be formulated as an inverse problem
\begin{equation}
	F(\mathbf{x}, \mathbf{c})\coloneqq(\mathcal{F}_S(\mathbf{x}\odot c_1), \cdots, \mathcal{F}_S(\mathbf{x}\odot c_{nc})) = \mathbf{y}~,
  \label{eq:model}
\end{equation}
where $\mathcal{F}_S$ is an undersampled multi-channel Fourier transform operator and the correspondingly obtained k-space data is $\mathbf{y} = (y_1, \cdots, y_{nc})^T$; $\mathbf{y}\in \mathbb{C}^{d\times nc}$, $\mathbf{x}\in \mathbb{C}^{n\times n}$ denotes the image content and $\mathbf{c}=(c_1, \cdots, c_{nc})^T$; $\mathbf{c}\in \mathbb{C}^{n\times n\times nc}$ denotes the coil sensitivities. $d$ is the number of samples in k-space, $nc$ is the number of coils, and $n\times n$ is the image size. 
\cref{eq:model} can be solved in the following two ways.

One common way for MR image reconstruction is to predetermine the coil sensitivities $\mathbf{c}$ from a reference
scan or from a fully sampled k-space center. Following coil estimation, we can solve the linear inverse problem
using by optimization regularized least-squares functional
\begin{equation}
  \underset{\mathbf{x}}{\min} \frac{1}{2}\|F_{\mathbf{c}}(\mathbf{x})-\mathbf{y}\|^2 + \alpha R(\mathbf{x}),
\end{equation}
where $F_{\mathbf{c}}(\mathbf{x}) := F(\mathbf{x}, \mathbf{c})$ is a linear operator and $R(\mathbf{x})$ is the
regularization term employing prior knowledge about the image, such as
$\ell^2$ regularization\cite{Ying_IEMBS_2004}, total variation\cite{Block_Magn.Reson.Med._2007}, $\ell^1$-sparsity\cite{Lustig_Magn.Reson.Med._2007},
or a learned log-likelihood function\cite{Luo_Magn.Reson.Med._2020}.

Alternatively, both image and coil sensitivities can be jointly estimated from the same acquired data\cite{Ying_Magn.Reson.Med._2007, Uecker_Magn.Reson.Med._2008}.
Ref. \cite{Uecker_Magn.Reson.Med._2008} formulates MR image reconstruction as a
nonlinear inverse problem and proposes to solve \cref{eq:model}
using the Iteratively Regularized Gauss Newton Method (IRGNM).
This method linearizes the nonlinear inverse problem at each Gauss Newton 
step $k$, and estimates the update $\delta \mathbf{m}\coloneqq(\delta \mathbf{x}, \delta \mathbf{c})$ 
for the pair $\mathbf{m}^k\coloneqq(\mathbf{x}^k, \mathbf{c}^k)$ by minimizing
a regularized least-squares functional for the linearized sub-problem 
\begin{align}
	\underset{\delta \mathbf{m}}{\min}&\frac{1}{2}\|F'(\mathbf{m}^k)\delta \mathbf{m}~+~F(\mathbf{m}^k)~-~\mathbf{y}\|^2\nonumber \\
  &~+~\beta^k\mathcal{W}(\mathbf{c}^k~+~\delta \mathbf{c})~+~\alpha^k{R}(\mathbf{x}^k~+~\delta \mathbf{x})~.
	\label{eq:sub}
\end{align}
Here, $\mathcal{W}(\mathbf{c})=\|w\odot\mathcal{F}\mathbf{c}\|^2$ is a penalty on the high Fourier coefficients
of the coil sensitivities and $R(\mathbf{x})$ is a regularization term on the image $\mathbf{x}$, e.g., $\ell_2$-norm\cite{Uecker_Magn.Reson.Med._2008},
$\ell_1$-sparsity in the wavelet domain\cite{Uecker__2008}, or
total variation\cite{Knoll_Magn.Reson.Med._2012}.
The $\alpha_k$ and $\beta_k$ are decreasing in each iteration step.

\subsection{Learned priors as regularization}

Learned prior knowledge can be used for regularization in image reconstruction.
Generative priors, such as variational autoencoder, autoregressive models (e.g.,
PixelCNN), and diffusion models, are used to incorporate empirical knowledge
about images into iterative optimization algorithms.  We want to use generative
priors directly as a drop-in replacement for conventional priors in existing
image reconstruction algorithms, which are often based on proximal methods.

The proximal operator for the log-prior $\log p(\mathbf{x})$ is defined as 
\begin{equation}
	\operatorname{prox}_t(\mathbf{z})~=~\underset{\mathbf{x}}{\arg \min}~\frac{1}{2t}\|\mathbf{x}-\mathbf{z}\|^2 + \log p(\mathbf{x})~.
\label{eq:prox}
\end{equation}
When the mapping above is not analytically computable, 
the proximal operator could be approximated by minimizing \cref{eq:prox} using gradient descent.
Note that the minimization problem for the proximal operator is the same as for a denoising problem for
complex Gaussian noise\cite{Romano_SIAMJ.Img.Sci._2017}.
Assuming some regularity of the prior and noise-like properties of the error
during reconstruction, the gradient can be expected to always point in the
same directions towards denoised images. Therefore, the optimality condition
at the solution is approximately
\begin{equation}
	0 \approx \frac{1}{t}( prox_t(\mathbf{z}) -  \mathbf{z} ) + \nabla_\mathbf{x} \log p(\mathbf{x}) ~.
\end{equation}
The solution is then equivalent to a single gradient-descent step with
an arbitrary initial guess and unit step size
\begin{align*}
	\operatorname{prox}_t(\mathbf{z}) &\approx  \mathbf{z} - t \nabla_\mathbf{x} \log p(\mathbf{x})~,
\end{align*}
which simply yields a gradient-descent step
for the log-prior in the overall algorithm.
In this work, two types of log-priors are used which are described below.

\begin{enumerate}[wide,labelindent=0pt]
\item[\textbf{PixelCNN prior:}]
This prior is formulated using a joint distribution over the elements of an image vector
\begin{equation}
  \log p(\mathbf{x};\mathrm{NET}(\hat{\Theta}, \mathbf{x}))=\log p(x^{(1)})\prod_{i=2}^{n^2} p(x^{(i)}\mid x^{(1)},..,x^{(i-1)})~,
  \label{eq:joint}
\end{equation}
where the neural network $\mathrm{NET}(\hat{\Theta}, \mathbf{x})$ predicts the distribution parameters of a mixture 
of logistic distributions that is used to describe every pixel
and where the dependencies between the channels for the real and imaginary
parts are described with nonlinear dependencies\cite{Luo_Magn.Reson.Med._2020}.
All these parameters used to probabilistically model the image are predicted by a causal network\cite{Salimans_ICLR_2017}
that encodes the relationship between pixels as formulated in \cref{eq:joint}. 
The gradient of $\log p(\mathbf{x};\mathrm{NET}(\hat{\Theta}, \mathbf{x}))$ with respect to $\mathbf{x}$ 
can be computed by back-propagation through the neural network. 

\item[\textbf{Probabilistic diffusion prior:}] The diffusion probabilistic model proposed in Ref. \cite{Sohldickstein_ICML_2015} is constructed with a forward Markovian process and a learned reverse process. The forward process is to gradually transfer
a data distribution $q(\mathbf{x}_0)$ to a smoother known distribution $q(\mathbf{x}_N)$, e.g., a Gaussian distribution, 
by adding noise to data points.
The reverse process is to undo this forward process with learned reverse transitions, which are described as
\begin{align}
	p_{\theta}(\mathbf{x}_{i-1}\mid \mathbf{x}_{i})&=\mathcal{CN}(\mathbf{x}_{i-1}\mid\mu_\theta(\mathbf{x}_{i},i),\tau_{i}^{2}\mathbf{I})\, ,
\label{eq:reverse}
\end{align}
where $\tau_i$
can be computed from the noise scales $\sigma_i$ at each step
and $\mu_\theta(\mathbf{x}_{i},i)$ can be understood as a denoised image
based on the smoothed prior at each noise level. Instead of learning this
distribution directly, the gradient of the log-prior is learned for all noise scales
\begin{align}
	\nabla_{\mathbf{x}_i}\log p_{\boldsymbol{\theta}}\left(\mathbf{x}_{i-1}\mid \mathbf{x}_{i}\right)
	& = \frac{1}{\tau_{i}^2}\left(\sigma_{i}^{2}-\sigma_{i-1}^{2}\right) \mathbf{s}_{\boldsymbol{\theta}}\left(\mathbf{x}_{i-1}, i\right)~,
\end{align}
where $\mathbf{s}_\theta(\mathbf{x}_i, i)$ is a trained score network\cite{Song_ICLR_2021},
which is computationally efficient because it avoids backpropagation.
We refer the reader to Ref. \cite{Luo_Magn.Reson.Med._2023} for details about this
method.
The reverse transitions start with $\sigma_{max}$ and end at a $\sigma_{min}\approx 0$. In this work we use $\sigma_{max}=0.3$ and $\sigma_{min}=0.01$, and 
\begin{equation}
	\sigma_i = \sigma_{min} + (\sigma_{max}-\sigma_{min})\cdot \log (1+(1-i/N)\cdot(\mathsf{e}-1))~.
  \label{eq:noise}
\end{equation}
Here, $\mathsf{e}$ is Euler constant, $i$ is bounded to the iteration step in the optimization algorithm 
, and $N$ is the number of noise scales which corresponds to the total number of iterations.
\end{enumerate}

For linear reconstruction, the two regularization terms can be directly plugged into 
proximal optimization algorithms available in image reconstruction frameworks,
such as the Fast Iterative Shrinkage-Thresholding Method (FISTA)\cite{Beck_SIAMJ.Img.Sci._2009}, or 
the Alternating Direction Methods of Multipliers (ADMM)\cite{Venkatakrishnan__2013}.

Similarly, for nonlinear inverse problems, we can apply the regularization terms to the linearized sub-problem in \cref{eq:sub}. In nonlinear reconstruction, the image content $\mathbf{x}$ is usually smooth at early Gauss Newton steps and the distribution of $\mathbf{x}$ is far from the learned empirical distribution. Correspondingly, in this work, the Gauss Newton optimization is split into two stages.
In the first stage, an $\ell_2$-norm regularization is applied, and the method of conjugate gradients (CG)
is used to minimize \cref{eq:sub}. In the second stage, i.e., the later Gauss-Newton steps, FISTA is utilized with the proximal operators. The entire algorithm is outlined in \cref{alg:seq}.

\begin{algorithm}[h]
	\caption{Two-stage IRGNM for NLINV reconstruction}
  \label{alg:seq}
	\begin{algorithmic}[1]
	\Inputs{y - kspace data, n - MaxIter, r - RegIter
	} 
	\Initialize{$\mathbf{x}^0=\mathbf{1},~\mathbf{c}^0=\mathbf{0},~\alpha^0=1, ~\beta^0=1$} 
	\For{$k~\mathrm{in}~\{0,\:\dots,\:\mathrm{n-1}\}$} 
		\If { $k< \mathrm{n - r }$}
      \State $R(\mathbf{x})~=~\|\mathbf{x}\|^2$
      \State Estimate $\delta \mathbf{x},~ \delta \mathbf{c}$ in \cref{eq:sub} with CG
    \Else
      \State $R(\mathbf{x})~=~\mathrm{log~p}(\mathbf{x})$\: or\: $\|\Psi \mathbf{x}\|^1$ 
      \State Estimate $\delta \mathbf{x},~ \delta \mathbf{c}$ in \cref{eq:sub} with FISTA
    \EndIf
    \State $\mathbf{x}^{k+1} = \mathbf{x}^k + \delta \mathbf{x},\: \mathbf{c}^{k+1} = \mathbf{c}^k + \delta \mathbf{c}$
    \State $\alpha^{k+1}=\max (\alpha_{\min },\: \alpha^k/2) $
    \State $\beta^{k+1}=\beta^k/2$
	\EndFor
	\end{algorithmic}
	
\end{algorithm}

\section{Methods}

In this section, we first describe how we implemented the proposed workflow, as shown in \cref{fig:overview},
for extracting prior knowledge from an image dataset and then how we use the learned prior
for regularization in image reconstruction. We detail how we evaluated the performance
of the priors used for image regularization in different settings.

\subsection{Preprocessing of the training dataset}

As for training data, we use human brain images from the Autism Brain Imaging Data Exchange (ABIDE)\cite{Martino_Mol.Psychiatr._2013,Martino_Scien.Data_2017}.
After we downloaded the dataset, which comes in 3D volumes in
NII\footnote[1]{The Neuroimaging Informatics Technology Initiative (NIfTI) is
an open file format commonly used to store brain imaging data obtained using
Magnetic Resonance Imaging methods.} format, we performed the following steps
to preprocess it.

\begin{enumerate}
  \item Load each 3D volume and resample it with the conform function of NiBabel\cite{Brett_Nibabel_5.1.0} to make its axial plane have a size of 256~$\times$~256. 
  \item Split the volume into 2D image slices that are oriented in axial plane.
  \item Add background Gaussian noise ($\mu=0.003, \sigma=5$) to all slices as it usually present in MR images, and then normalize every slice by dividing its maximum pixel value.
  \item Crop a 30~$\times$~30 patch from the XY-corner of each normalized slice and compute the mean $\mu$ and standard deviation $\sigma$ over all pixels of the patch.
  \item Exclude slices (without image content) from phase augmentation when mean $\mu < 0.04$ and standard deviation $\sigma < 0.0061$.
\end{enumerate}

\subsection{Phase augmentation}

ABIDE images are provided solely as magnitude images without phase information. 
The magnitude of an MR image is determined by the proton density, relaxation effects,
and receive fields, while the phase is affected by the phase of the receive field, inhomogeneities
of the static field, eddy currents, and chemical shift. Phase augmentation can be used
to add a phase to obtain more realistic complex-valued images. Here, we describe 
a procedure to obtain new samples with phase information from magnitude images
using a prior previously trained prior for complex-valued images.
The method is based on previous research\cite{Sohldickstein_ICML_2015} for the sampling
of a posterior. Given the likelihood term of the magnitude $p(\mathbf{m}|\mathbf{x})$ and a prior for
complex-valued images $p(\mathbf{x})$, the posterior of the complex image is proportional to
$p(\mathbf{x}|\mathbf{m}) \propto p(\mathbf{x}) \cdot p(\mathbf{m}|\mathbf{x})$ where
$$\bar{\mathbf{m}}=\sqrt{\mathbf{x}_r^2 + \mathbf{x}_i^2 }\ ,$$
and $\mathbf{x}_r$ and $\mathbf{x}_i$ are the real part and imaginary part, respectively.
The likelihood term for a given magnitude image $\mathbf{m}$ is $p(\mathbf{m}|\mathbf{x}) = \delta(\mathbf{m}-\sqrt{\mathbf{x}_r^2 + \mathbf{x}_i^2})$.
To be able to apply gradient-based methods, we approximate this with a narrow Gaussian distribution
\begin{equation}
  p(\mathbf{m}|\mathbf{x}) \propto \exp\bigl(-\epsilon\bigl\|\mathbf{m}-\sqrt{\mathbf{x}_r^2 + \mathbf{x}_i^2 }\bigr\|_2^2\bigr) .
\end{equation}

Specifically, we initialize samples with random complex Gaussian noise
and then transfer them gradually to the distribution of complex images
with learned transition kernels ${p}_\theta(\mathbf{x}_{n}|\mathbf{x}_{n+1})$. 
We run unadjusted Langevin iterations sequentially at each intermediate distribution
\begin{equation}
  \mathbf{x}_n^{k+1} \leftarrow \mathbf{x}_n^{k} + \frac{\gamma}{2}\nabla_{\mathbf{x}}\log {p}_\theta(\mathbf{x}_{n}^{k}\mid\mathbf{x}_{n+1}^\mathrm{K})+\frac{\gamma
  }{2}\nabla_{\mathbf{x}}\log p(\mathbf{m}|\mathbf{x}_n^k) + \sqrt{\gamma}\mathbf{z}~.
  \label{eq:ld}
\end{equation}
Here, $\mathbf{z}$ is complex Gaussian noise, which introduces random
fluctuations and $\gamma$ controls the step size of the Langevin algorithm.
The sampling algorithm was implemented with TensorFlow and used with the 
pre-trained generative model $\mathrm{NET}_1$ from Ref. \cite{Luo_Magn.Reson.Med._2023},
which was trained on the small dataset from Ref. \cite{Luo_Magn.Reson.Med._2020}.
For each magnitude image five complex images were generated.

\subsection{Training of priors}

\begin{table*}
  \caption{Datasets and computational resources used to train the six different priors used in this work. \texttt{P}\textsubscript{SC} trained from
a small dataset of complex-valued images will serve as
a baseline for the other learned priors.\label{tab:priors}}
  \centering
  \resizebox{\textwidth}{!}{%
  \begin{tabular}{c@{\hskip 10pt} c @{\hskip 10pt}c@{\hskip 10pt}c @{\hskip 10pt}c @{\hskip 10pt}c@{\hskip 10pt}c @{\hskip 10pt}c @{\hskip 10pt}c}%
    \toprule
	  Prior & &  Model & Phase & Nr. of Images & MR Contrasts & GPUs & Parameters & Time~$\times$~epochs \\ \hline
	  \texttt{P}\textsubscript{SC} & (small, complex)& PixelCNN & preserved & 1000   &  T\textsubscript{1}, T\textsubscript{2}, T\textsubscript{2}-FLAIR, T$^*_\mathrm{2}$ &4$\times$A100, 80G & \textasciitilde22M & \textasciitilde40s~$\times$~500 \\
	  \texttt{P}\textsubscript{SM} & (small, magnitude)& PixelCNN & not available & 1000     &  T\textsubscript{1}, T\textsubscript{2}, T\textsubscript{2}-FLAIR, T$^*_\mathrm{2}$ &4$\times$V100, 32G & \textasciitilde22M & \textasciitilde144s~$\times$~500 \\
	  \texttt{P}\textsubscript{LM} & (large, magnitude)& PixelCNN & not available & 23078    & MPRAGE &4$\times$A100, 80G & \textasciitilde22M & \textasciitilde748s~$\times$~100 \\
	  \texttt{P}\textsubscript{LC} & (large, complex) & PixelCNN & generated & 23078  & MPRAGE &3$\times$A100, 80G & \textasciitilde22M & \textasciitilde1058s~$\times$~100 \\
	  \texttt{D}\textsubscript{SC} & (SMLD, complex) & Diffusion & generated & 79058 & MPRAGE & 4$\times$A100, 80G & \textasciitilde8M &\textasciitilde2330s~$\times$~50 \\
\bottomrule 
\end{tabular}
  }
\end{table*}

In total, we trained six priors in this work.
The PixelCNN priors, \texttt{P}\textsubscript{SC} and \texttt{P}\textsubscript{SM}, were trained on the
small brain image dataset used in Ref. \cite{Luo_Magn.Reson.Med._2020} using complex and magnitude images, respectively.
\texttt{P}\textsubscript{LM} and \texttt{P}\textsubscript{LC} were trained on a subset of the preprocessed ABIDE dataset
corresponding of 500 volumes and the corresponding phase-augmented complex images, respectively.
We also trained one diffusion priors, SMLD (\texttt{D}\textsubscript{SC}) with phase-augmented images using the full ABIDE dataset with 1206 volumes.

During the training, images were normalized to have a maximum magnitude of one and then subjected
to random mirroring, flipping, and rotation prior to being fed into the neural network. 
Complex images were fed as two-channel maps (i.e., real and imaginary), and magnitude images were
fed as single-channel maps. 
The networks used for PixelCNN and the diffusion models were implemented with TensorFlow (TF)
and the optimizer ADAM was used for all training tasks, which was performed using multi-GPU
systems using different GPUs (Nvidia Corporation, Santa Clara, CA, USA).
More information about priors and training is detailed in \cref{tab:priors}. 

\begin{enumerate}[wide,labelindent=0pt]
  \item[\textbf{PixelCNN prior:}] We trained this generative model by maximizing the probability of the joint distribution
over all the pixels in the image using the discretized logistic mixture distribution loss proposed in 
Ref. \cite{Salimans_ICLR_2017}.

  \item[\textbf{Diffusion prior:}] 
There exist two types of diffusion models that are based on the denoising score matching method\cite{Vincent_NeuralComput._2011}, namely, denoising Score
Matching with Langevin Dynamics (SMLD)\cite{Song_NIPS_2019} and Denoising Diffusion Probabilistic Models (DDPM)\cite{Ho_NIPS_2020}. Both are
unified in a common framework described in Ref. \cite{Song_ICLR_2021}.
We train diffusion priors using SMLD with the same Refine-Net\cite{Lin_arXiv_2016} architecture also used in Ref. \cite{Luo_Magn.Reson.Med._2023}.
The loss function used to train the score network $\mathbf{s}_{\theta}(\mathbf{x}_i,i)$ is given by
  \begin{equation}
    \theta^{\ast}=\underset{\theta}{\arg\operatorname*{min}}~\sum_i\mathbb{E}_{\mathbf{x}_{0}}\mathbb{E}_{\mathbf{x}_{i}|\mathbf{x}_{0}}\bigl[\lambda_{i}\bigl\|\mathbf{s}_{\theta}(\mathbf{x}_{i},i)-\nabla_{\mathbf{x}_{i}}\log p\left(\mathbf{x}_{i}\mid\mathbf{x}_{0}\right)\bigr\|_2^{2}\bigr]\,
    \label{eq:loss}
  \end{equation}
where $\lambda_{i}$ is the weighting function described in Ref. \cite{Song_ICLR_2021}. 
\end{enumerate}

\subsection{Experimental evaluation}

In this section, we use Berkeley Advanced Reconstruction Toolbox (BART)\cite{Uecker__2015} to evaluate the trained priors using
Parallel Imaging Compressed Sensing (PICS) and Nonlinear Inversion (NLINV). 
The corresponding commands have an option for loading an exported TensorFlow computation graph and using it for regularization.
The exported graph was wrapped into a nonlinear operator and then used in the proximal mapping step\cite{Blumenthal_Magn.Reson.Med._2023}.
For the linear reconstruction using PICS, the coil sensitivities are estimated with ESPIRiT\cite{Uecker_Magn.Reson.Med._2014}.
In the nonlinear reconstruction using NLINV, \cref{alg:seq} was implemented with the nonlinear operator framework\cite{Blumenthal_Magn.Reson.Med._2023}.

We then performed different experiments using PICS and NLINV for the six priors
using different sampling patterns in comparison to zero-filled, $\ell_2$, $\ell_1$-wavelet
reconstructions and coil-combined images form fully sampled data. Here, we used
a T\textsubscript{1}-weighted k-space (a common MR image contrast) from the test dataset used in Ref. \cite{Luo_Magn.Reson.Med._2020}.
We additionally performed a study using quantitative image quality metrics using 
3D MPRAGE data to evaluate the impact of the size of the training dataset,
and performed an evaluation study with human readers for six fully sampled
3D TurboFLASH datasets as described below.

\begin{enumerate}[wide,labelindent=0pt]
  \item[\textbf{The influence of phase maps:}] We performed retrospective reconstruction using all six priors. 
Three types of undersampling pattern were used in this retrospective experiment, including five-fold acceleration along phase direction, 
two-times and three times acceleration along frequency and phase direction, respectively, and 8.2-times undersampling using Poisson-disc sampling.
While the acceleration along frequency-encoding direction is not realistic, we use it to explore how the priors handle different 2D different 
patterns. In a later experiment below, we then use 3D k-space acquisitions where these sampling pattern are feasible.
  \item[\textbf{The influence of the size of dataset:}] We performed the reconstruction using \texttt{P}\textsubscript{SC} and \texttt{P}\textsubscript{LC}.
The k-space data were acquired from the brain of a healthy volunteer using MPRAGE sequence on 3T Siemens Skyra scanner
(Siemens Healthineers, Erlangen, Germany) with 16-channel head coils. The protocol parameters were:
TE~=~2.45~ms, flip angle $\alpha=8$\textdegree, TI~=~900~ms, and TR~=~2000~ms, 4/5 partial parallel Fourier imaging, and 2-fold acceleration along one
phase encoding direction. This acquired 3D volume has dimensions 256 $\times$ 256 $\times$ 224 and isotropic voxel size of 1~mm.
We further undersampled the acquired 3D k-space data two and three times along two phase-encoding directions with the central
region of size 30~$\times$~25 reserved. The reconstruction was performed slice-by-slice in a 2D plane.
To quantitatively assess the robustness of the priors, we computed PSNR and SSIM for the \texttt{P}\textsubscript{LC}-regularized reconstruction 
from 4 and 6-times undersampled k-space against the \texttt{P}\textsubscript{LC}-regularized reconstruction from 2-times undersampled k-space 
and compared it to PSNR and SSIM for the $\ell_2$-wavelet regularization and the prior \texttt{P}\textsubscript{SC}. 
\item[\textbf{3D reconstruction quality:}] For reconstruction of 3D data sets, we chose the 
diffusion models, \texttt{D}\textsubscript{SC}, which is less computationally expensive than PixelCNN. 
The k-space data was acquired from 6 volunteers using a 3D TurboFLASH sequence (TE~=~3.3~ms, TR~=~2250~ms, TI~=~900~ms, flip angle $\alpha=9$\textdegree)
using a 3T Siemens Prisma scanner and a 64-channel head-coil.
These acquired 3D volumes have dimensions 256~$\times$~256~$\times$ 176~and isotropic size of 1~mm.
We undersampled the acquired 3D k-space data using a prospectively feasible Poisson-disc 
pattern with 8.2-time undersampling. 
When applying the prior during reconstruction, all slices in the axial plane are padded to a size of 256~$\times$~256
Following this, the prior is applied on all slices in parallel and the computed gradient will be
resized to the original image size. The reconstructed volumes were blindly evaluated 
by three clinicians with variable experience in neuroimaging (\textasciitilde20 years, \textasciitilde10 years, \textasciitilde5 years).
$\ell_1$-wavelet reconstructions and a reference reconstructed from fully sampled k-space data by coil-combination
were included in this evaluation study.
The grading scale used in this study ranged from 5 to 1, where a score of 5 represents "excellent" image quality and
a 1 denotes "bad" image quality.
\end{enumerate}

\section{Results}

\begin{figure}
  \centering
  
      \includegraphics[width=\textwidth]{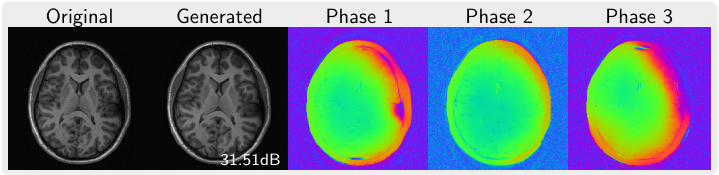}
  
\caption{Human brain images. On the left, the original magnitude-only image is
compared to the magnitude of a corresponding image generated using phase augmentation. On the right, the
phase maps of three different generated images are shown. \label{fig:1}}
\end{figure}

\cref{fig:1} shows an example of a magnitude image from the ABIDE dataset
and the corresponding magnitude and phase maps of
complex-valued images generated in the stage of phase augmentation. The magnitude part
of the generated images stays very close to the original
image but exhibits a bit less noise. The phase of
the generated images maps is smooth and looks realistic
with some random variations introduced by the sampling algorithm.



\subsection{The influence of phase maps}

\begin{figure}
  \centering
  \includegraphics[width=\textwidth]{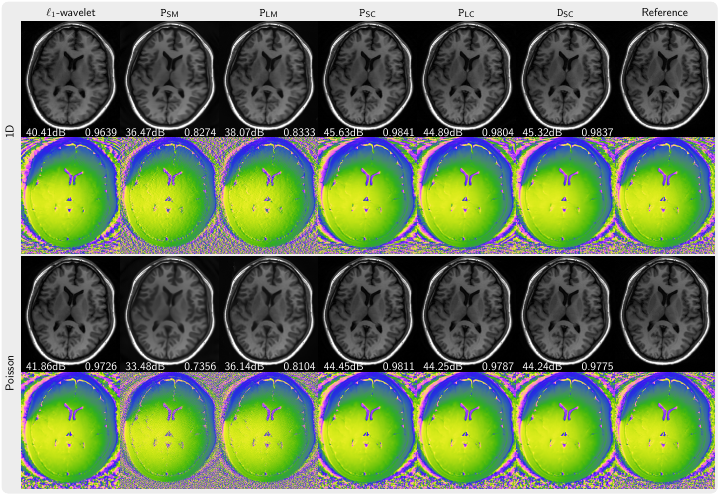}
	\caption{Comparison of images reconstructed using PICS using the priors \texttt{P}\textsubscript{SM}, \texttt{P}\textsubscript{LM}, \texttt{P}\textsubscript{SC}, \texttt{P}\textsubscript{LC}, \texttt{D}\textsubscript{SC} in comparison to an $\ell_1$-wavelet reconstruction and a reference (c.f. error maps in the supplementary). The top two rows (1D) present the results for 5-fold acceleration along phase-encoding direction with 30 calibration lines. The bottom two rows (Poisson) show the results using a Poisson-disc acquisition of 8.2x-undersampling. PSNR and SSIM values are shown in white text.\label{fig:pics}}
\end{figure}

\begin{figure}
  \centering
  \includegraphics[width=\textwidth]{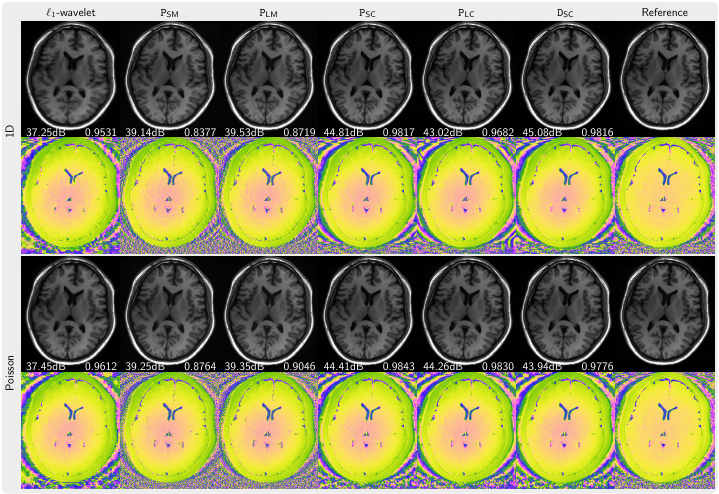}
	\caption{Comparison of images reconstructed using NLINV using the priors \texttt{P}\textsubscript{SM}, \texttt{P}\textsubscript{LM}, \texttt{P}\textsubscript{SC}, \texttt{P}\textsubscript{LC}, \texttt{D}\textsubscript{SC} in comparison to an $\ell_1$-wavelet reconstruction
	and a reference (c.f. error maps in the supplementary). The top two rows (1D) present the results for 5-fold acceleration along phase-encoding direction with 30 calibration lines.
	The bottom two rows (Poisson) show the results using a Poisson-disc acquisition of 8.2x-undersampling. PSNR and SSIM values are shown in white text.\label{fig:nlinv}}
\end{figure}

\begin{figure}
	\centering
	\includegraphics[width=\textwidth]{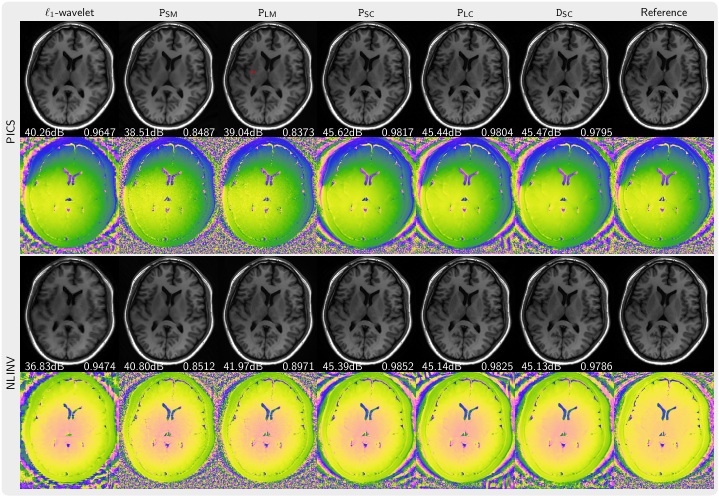}
	\caption{Comparison of images reconstructed using NLINV and PICS using the priors \texttt{P}\textsubscript{SM}, \texttt{P}\textsubscript{LM}, \texttt{P}\textsubscript{SC}, \texttt{P}\textsubscript{LC}, \texttt{D}\textsubscript{SC} for a 2~$\times$~3 sampling pattern in comparison to an $\ell_1$-wavelet reconstruction and a reference (c.f. error maps in the supplementary). PSNR and SSIM values are shown in white text. Artifacts (red arrow) are introduced by the priors trained on magnitude images when using PICS.}
  \label{fig:mag_vs_cplx}
\end{figure}

\cref{fig:pics} presents the magnitude and phase of images that are reconstructed
using PICS with priors trained from magnitude image, complex images with
preserved phase, and complex images with generated using our phase augmentation method
While the priors \texttt{P}\textsubscript{SM} and \texttt{P}\textsubscript{LM}
trained from magnitude images can remove folding artifacts introduced by
undersampling, they exhibit over-smoothing of the magnitude as indicated
by its lower PSNR and SSIM values and also demonstrates poor capabilities
in denoising the phase.
In contrast, the prior \texttt{P}\textsubscript{SC} trained on complex-valued
images performs much better. Furthermore, the priors \texttt{P}\textsubscript{LC}
and \texttt{D}\textsubscript{SC} trained on phase-augmented images and 
perform almost as well.  Very similar results were obtained for NLINV
as shown in \cref{fig:nlinv}.
In \cref{fig:mag_vs_cplx}, the k-space is sampled using 2~$\times$~3 pattern.
We observed artifacts (red arrow) introduced by the priors trained from
magnitude-only images reconstructed with PICS method, but not with NLINV method.
Under all investigated conditions, the priors trained on complex-valued images 
outperform the reconstruction with $\ell_1$-wavelet regularization.

\subsection{The influence of the size of dataset}

\begin{figure}
	\centering
	\includegraphics[width=\textwidth]{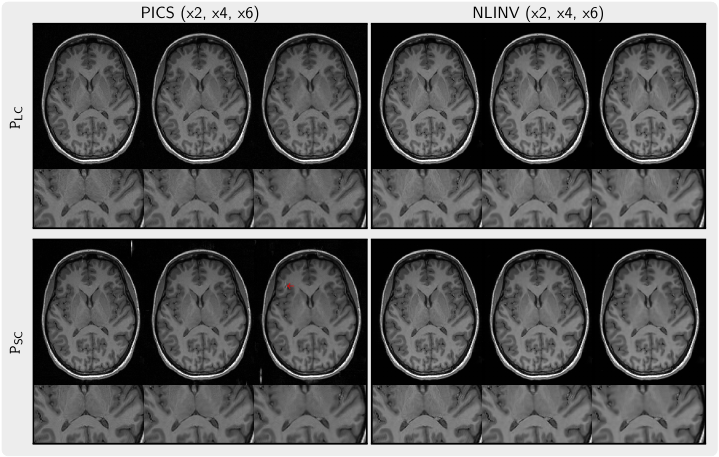}
	\caption{Comparison of images reconstructed with PICS (left) and NLINV (right) using priors \texttt{P}\textsubscript{LC} (top) and \texttt{P}\textsubscript{SC} (bottom) trained on small and large datasets. We observed artifacts (red arrow) when using PICS with the prior  \texttt{P}\textsubscript{SC} trained on the small dataset. The images in each column are reconstructed from k-space undersampled with factors ranging from 2 to 6 (left to right).}
  \label{fig:size}
\end{figure}

\cref{fig:size} presents the images regularized by the priors (\texttt{P}\textsubscript{SC} and \texttt{P}\textsubscript{LC}) trained on small and large datasets, respectively.
When using PICS with the prior \texttt{P}\textsubscript{SC} artifacts become apparent in the background and in the brain, whereas no such artifacts are
observed when applying the prior \texttt{P}\textsubscript{LC}.  Furthermore, image details appear to be better preserved with high undersampling 
for the prior \texttt{P}\textsubscript{LC}. 

\begin{figure}
	\centering
	\includegraphics[width=0.95\textwidth]{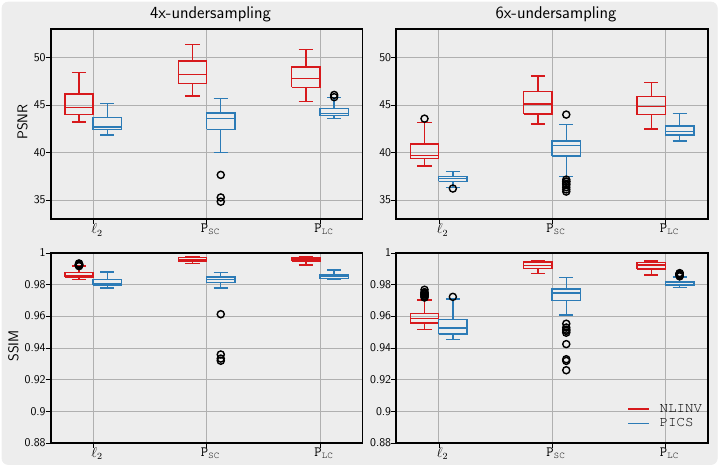}
	\caption{PSNR (top) and SSIM (bottom) metrics for images reconstructed with PICS (blue) and NLINV (red) when using $\ell_2$-regularization and when using PixelCNN priors \texttt{P}\textsubscript{SC} and \texttt{P}\textsubscript{LC} for 4x (left) and 6x (right) undersampling relative to a reconstruction from 2x undersampled k-space and using \texttt{P}\textsubscript{LC}. The PixelCNN trained on a larger dataset, \texttt{P}\textsubscript{LC}, shows fewer outliers compared to the one trained on a smaller dataset, \texttt{P}\textsubscript{SC}.
	\label{fig:boxplot}}
\end{figure}

These observations can be confirmed quantitatively.
We display the three sets of PSNR and SSIM metrics for $\ell_2$, \texttt{P}\textsubscript{SC}, and \texttt{P}\textsubscript{LC} with boxplots in \cref{fig:boxplot}
for 4x and 6x undersampling relative to a reconstruction from 2x undersampled k-space and using \texttt{P}\textsubscript{LC}.
Here, the $\ell_2$-regularization serves as a baseline reconstruction which is not influenced by the properties of a learned prior.
For the \texttt{P}\textsubscript{LC} prior learned from a large dataset there are only a few outliers above the average when using PICS and NLINV.
However, when using the \texttt{P}\textsubscript{SC} prior learned from a small dataset there are many outliers below the average,
especially when  the undersampling factor is high in the case of PICS.

\subsection{3D reconstruction using diffusion priors}

\begin{figure*}
	\centering
	\includegraphics[width=\textwidth]{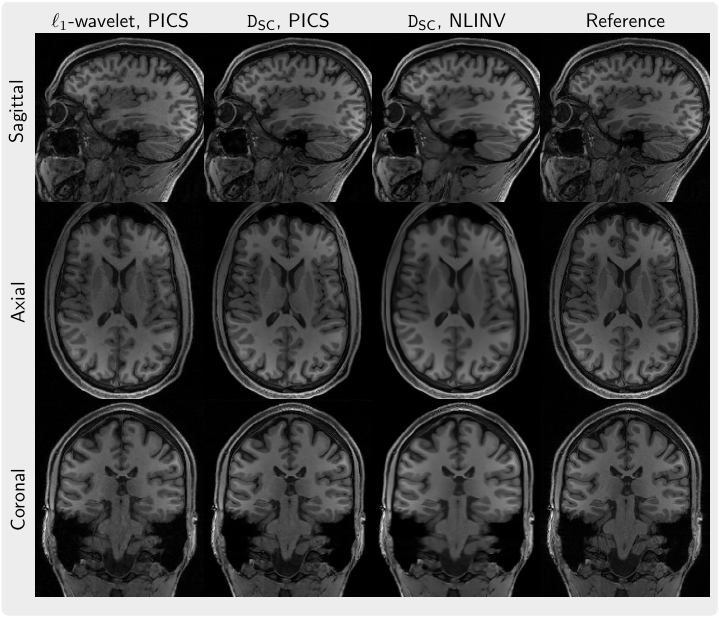}
	\caption{Slices in three orientations (Sagittal, Axial, Coronal, from top to bottom) from a 3D volume reconstructed using
	PICS from 8.2x-undersampled k-space with Poisson-disc sampling with $\ell_1$-wavelet regularization and the diffusion prior \texttt{D}\textsubscript{SC}
	and using NLINV with diffusion prior \texttt{D}\textsubscript{SC} (from left to right).}
  \label{fig:3drecon}
\end{figure*}

As an example, \cref{fig:3drecon} presents three slices in the sagittal, axial, and coronal planes 
for a 3D volume reconstructed using the diffusion prior \texttt{D}\textsubscript{SC}
using PICS and NLINV in comparison to $\ell_1$-wavelet regularization and a reconstruction
by coil combination of Fourier-transformed fully-sampled k-space data.
By visual inspection, the $\ell_1$-regularized images appear to have reduced sharpness compared 
to the images regularized by the diffusion prior \texttt{D}\textsubscript{SC} while also having 
more noise.

\begin{figure*}
	\centering
	\includegraphics[width=0.6\columnwidth]{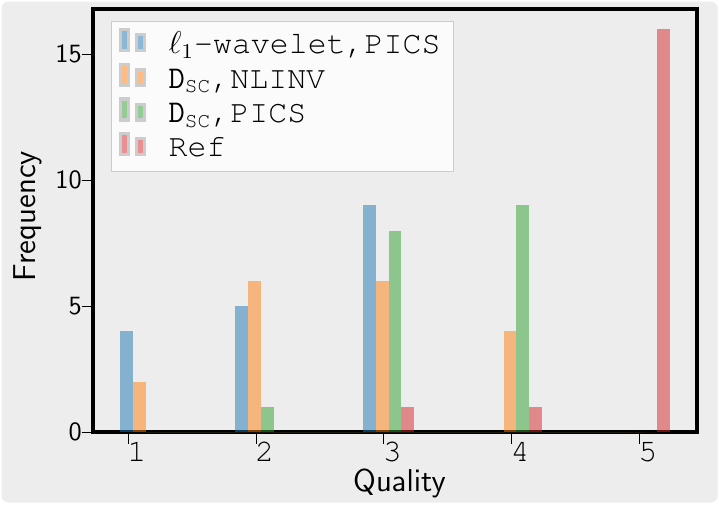}
	\caption{Blind evaluation by three clinicians of the volumes that are reconstructed with PICS using $\ell_1$-wavelet regularization, PICS with diffusion prior \texttt{D}\textsubscript{SC}, NLINV with the diffusion prior, and using coil combination of Fourier-transformed fully-sampled k-space data. The grading scale ranges from 5 (excellent) to 1 (bad). There are 18 numbers
  indicating the image quality for each method, and the y-label "Frequency" means how often
  a certain scale of image quality appears in the evaluation.}
  \label{fig:eval}
\end{figure*}

\cref{fig:eval} shows the results from the evaluation by clinical readers.
The diffusion prior \texttt{D}\textsubscript{SC} outperforms $\ell_1$-wavelet
regularization leveraging the learned knowledge. Here, \texttt{D}\textsubscript{SC} demonstrates
better performance when using PICS method compared to NLINV method.  With the
relatively high acceleration factor of 8.2 used, none of the reconstructions
matches the quality of the reference.

\section{Discussion}

A practical workflow was presented for extracting prior information from a set of magnitude-only images. 
It starts with the preparation of the training dataset, then followed by the generative modelling of
complex-valued images and ends with the application of generative priors for regularization in
image reconstruction. The effectiveness of the prior in boosting image quality was assessed
by clinicians. Different aspects of this work are discussed below.

\noindent
To exploit the information in magnitude-only images, the prediction of
phase maps using an U-net was reported in Ref. \cite{Luo__2022a}. Different to that, we prepared
a training dataset of phase-augmented images through conditional generation using a complex diffusion
prior that is first trained on a small dataset of complex-valued images. The prior trained
on a large dataset of phase-augmented images exhibits high robustness, as shown in \cref{fig:boxplot}.
The proposed approach will allow us to leverage the information in the large number of DICOM
images already available in the archives of radiology departments.

In Ref. \cite{Deveshwar_MDPI_BIOENG_2023}, the authors applied a conditional
generative adversarial network (GAN) to produce phase maps based on magnitude images
and then used them to synthesize k-space data. They reported the comparable
performance to raw k-space data when the synthetic k-space data 
were utilized to train a variational network\cite{Hammernik_Magn.Reson.Med._2017}
for reconstruction. However, this still required running ESPIRiT on prior
ground truth data from fastMRI\cite{Zbontar_arXiv_2019} to obtain
sensitivity maps for simulations.
Ref. \cite{Zijlstra__2023} expanded this idea by generating coil sensitivity phase maps based on 
magnitude images. This was achieved through a three-stage approach, involving the generation of 
low-resolution coil sensitivity phase maps based on magnitude images with a GAN,
upsampling of low-resolution maps to high-resolution ones,
and transformation of the coil images to k-space data.
Our work proposes a simple and less computationally intensive approach
based on phase augmentation using a generic diffusion prior trained
on complex-valued images.  The advantage of this framework
is that the learned prior is  independent of k-space sampling patterns and coil sensitivities,
and that it can be used as a regularization term in conventional
reconstruction algorithms.

\noindent 
Training a prior is computationally expensive. For example, it took
around 18 minutes to train \texttt{P}\textsubscript{LC} per epoch
through data parallelism using three A100 80G GPUs
(Nvidia Corporation, Santa Clara, CA, USA). In contrast, use
of the prior in conventional reconstruction algorithms is
computationally efficient. While previous reports indicate that
up to 2000 evaluations of a diffusion prior are needed for
reconstructing a single image\cite{Chung_Med.Image.Anal._2022}, 
the number of evaluations required for a conventional linear
reconstruction algorithm as used in this work is only about 100. This work demonstrates that how the curation of dataset can significantly influence the performance and robustness of the prior in image reconstruction. The more systematical comparison of different generative models would be the future work where the focus will be on the computational efficiency, robustness, and reconstruction quality.

\section{Conclusion}

This work focuses on how to extract prior knowledge from existing
magnitude-only image datasets using phase augmentation with generative models.
The extracted prior knowledge is then applied as regularization in image
reconstruction. The effectiveness of this approach in improving image quality
is systematically evaluated  across different settings. Our findings stress the
importance of incorporating phase information and leveraging large datasets to
raise performance and reliability of generative priors for MRI reconstruction.

\section{Data Availability}

In the spirit of reproducible research, the code used to preprocess the dataset
and to generate complex-valued images by phase augmentation using the diffusion prior,
and the shell scripts for the reconstruction are made available in our
repository: \url{https://github.com/mrirecon/image-priors}.
The Python library \texttt{spreco} used to train priors is available in this repository: {\url{https://github.com/mrirecon/spreco}}.
Pre-trained models are made available at Zenodo: \url{https://doi.org/10.5281/zenodo.8083750
}.
We refer readers to the webpage of our repository for additional information on the released materials.

\section{Acknowledgement}
We acknowledge funding by the "Niedersächsisches
Vorab" funding line of the Volkswagen Foundation.
Supported by the DZHK (German Centre for Cardiovascular Research), funding code: 81Z0300115

\bibliographystyle{ieeetr}
\bibliography{radiology.bib}
\end{document}